\documentclass[runningheads]{llncs}
\usepackage{epsfig}

\usepackage{microtype}
\usepackage{kantlipsum}
\usepackage{url}

\usepackage{graphicx}

\usepackage{multicol}
\usepackage{pslatex}
\usepackage[utf8]{inputenc}

\usepackage{multirow}
\usepackage{fancyhdr}
\usepackage{epic}
\usepackage{eepic}
\usepackage{amsmath}
\usepackage{amscd}
\usepackage{here}
\usepackage{lscape}
\usepackage{longtable}
\usepackage{comment}

\usepackage{url}
\usepackage{amsmath} 
\usepackage{float}
\usepackage[caption = false]{subfig}
\microtypesetup{activate=true}

\begin{document}

\title{Empirical Security and Privacy Analysis of Mobile Symptom Checking Applications on Google Play}

\author{I Wayan Budi Sentana \and Muhammad Ikram \and Mohamed Ali Kaafar \and Shlomo Berkovsky}
\authorrunning{Sentana et al.}

\institute{
{Macquarie University}\\
\email{i-wayan-budi.sentana@hdr.mq.edu.au, \{muhammad.ikram, Dali.Kaafar, Shlomo.Berkovsky\}@mq.edu.au}
}


\maketitle

\begin{abstract}
Smartphone technology has drastically improved over the past decade.  These improvements have seen the creation of specialized health applications, which offer consumers a range of health-related activities such as tracking and checking symptoms of health conditions or diseases through their smartphones.  We term these applications as \emph{Symptom Checking} apps or simply \emph{SymptomCheckers}. Due to the sensitive nature of the private data they collect, store and manage, leakage of user information could result in significant consequences.  In this paper, we use a combination of techniques from both static and dynamic analysis to  detect,  trace  and  categorize  security and privacy issues in 36 popular SymptomCheckers on Google Play. Our analyses reveal that SymptomCheckers request a significantly higher number of sensitive permissions and embed a higher number of third-party tracking libraries for  targeted advertisements and analytics exploiting the privileged access of the SymptomCheckers in which they exist, as a mean of collecting and sharing critically sensitive data about the user and their device. We find that these are sharing the data that they collect through unencrypted plain text to the third-party advertisers and, in some cases, to malicious domains. The results reveal that the exploitation of SymptomCheckers is present in popular apps, still readily available on Google Play.
%
\end{abstract}

Smartphone users are increasingly using their smartphones for health-related activities. The majority of smartphones now have the ability to passively collect health data as users progress through their day \cite{trifan2019passive}. The result of this passive logging has been the creation of specialized health-related mobile applications (SymptomCheckers) which track, manage and store the health data of users. The logging capabilities of SymptomCheckers go beyond passive tracking, as users can self-monitor their activities and manually record their personal data. A wide variety of categories are offered including; exercise and fitness tracker, sleep patterns and quality, cardiology and vascular health, mental and emotional health, and blood sugar levels for diabetes~\cite{smahel2019functions}.

Leveraging upon previous work~\cite{ikram_vtscore}, this paper presents the first characterization study of SymptomCheckers with a focus on security and privacy offered by these apps. In particular, we perform static and dynamic analysis to analyze the Android permissions, the presence of malicious code as well as third-party tracking libraries, and investigate the (un)encrypted traffic for content and end-points of the exfiltrated sensitive data. 



We collect and extract from a corpus of more than 1.7
million Android apps, 36 SymptomChecker for which the name or
the description suggest they enable to either track or check health-related activities of users. We then manually check that the apps actually
fall into the category of SymptomChecker (\S~\ref{data_collection}).
We use a set of tools to decompile the SymptomChekers and
analyze the source code of each of the mobile SymptomChekers.
We then inspect the apps to reveal the presence of third-party tracking libraries and sensitive permissions for critical resources on users' mobile devices. We summarize our analysis as follow: 
\begin{itemize}
    \item 13.8\% (5) of SymptomCheckers' use a vulnerable encryption scheme (i.e., SHA1 + RSA) for signing certificates and security purposes. 
    \item  44.4\% (16), 58.3\% (21), and 63.86\% (23) of SymptomCheckers provide Exported Activities, Exported Services and Exported Broadcast Receiver, respectively, which can be exploited by a malicious app. 
    We found that 10.81\% (4) of SymptomCheckers contain malware code embedded in their source codes. 
    \item To avoid source code analysis 89\% (32) of SymptomChekers rely upon different types of anti-analytics or obfuscation techniques. 
    \item 22.2\% (8) of SymptomChekers embed at least five different third-party tracking and advertisement libraries sharing sensitive user information such as location with third-party analytics and advertisers. 
    \item 5\% of the traffic generated by SymptomCheckers' use insecure HTTP protocol for transmitting users' sensitive data in plaintext which can be intercepted and modified by malicious in-path proxies. 
\end{itemize}

\section{\uppercase {Data Collection and Analysis Methodology}}

In this Section, we present our data collection and analysis methodology. 
\subsection{Data Collection Methodology}
\label{data_collection}
Given that Google Play store does not contain ``Symptom Checking'' apps' category, we devise a search methodology to find SymptomCheckers 
on Google Play. First, we 
use several keywords including ``symptom'', ``SymptomChecker'', and ``health checker'' in the description of 1.7 million Android applications collected in ~\cite{AdBlocking}. We obtained 353 apps that match those keywords. 
We conduct manual check apps description to ensure our SymptomCheckers apps are real {\it SymptomChecker}. As the research scope is to analyze the SymptomCheckers apps that are intended for end-user, we removed apps that only containing symptom trackers, health dictionaries or apps that are used to store \emph{only} user health history. We also discard the apps used for the learning process by medical students or the apps used to assist health practitioners. Overall we found 36 apps that met our search criteria. 

We then use \texttt{gplaycli}~\cite{gplaycli} to download 36 apps. 
We then installed these four apps on a mobile device via Android Debug Bridge (ADB) {\tt shell}, a development tool that facilitates communication between an Android device and a personal computer, for further analysis. 
We also scraped textual information including apps unique identifier, category and price,  regional availability, description, number of installs, developer information, user reviews, and apps rating. 
%
%

Table \ref{tab:top 10 apps by no of install} shows the top 10 SymptomCheckers apps sort by number of install and average ratings. Among the 36 apps, 83.3\% (30) apps have at least 1,000 devices. We found that WebMD~\cite{WebMD} is the most popular app with at least 10 Millions installs. The average rating, number of raters and number of reviews are respectively 3.27, 10646.81, and 4341.35, showing the popularity of SymptomCheckers among users. We also found that ADA~\cite{Ada} is highly rated by 290,484 users with an average rating of 4.74. 

\begin{table}
\centering
  \caption{Top 10 Free SymptomCheckers sort by number of installs. The lower part of the table summarises the average statics of SymptomCheckers.}
  \label{tab:top 10 apps by no of install}


  \begin{tabular}{|l|l|r|r|}
    \hline\hline
        {\bf \# }& {\bf Apps Name}  & {\bf \# of Installs} & {\bf Rating} \\
   \hline
        1 & com.webmd.android & 10,000,000+ & 4.44 \\
        2 & com.ada.app & 5,000,000+ & 4.74 \\
        3 & md.your & 1,000,000+ & 4.1\\
        4 & com.mayoclinic.patient & 1,000,000+ & 3.9 \\
        5 & com.programming. & 500,000+ & 4.52 \\
         & progressive.diagnoseapp & &  \\
        6 & com.symptomate.mobile & 100,000+ & 4.41 \\
        7 & nl.japps.android.depressiontest & 100,000+ & 3.82 \\
        8 & air.com.sensely.asknhs & 100,000+ & 4.29 \\
        9 & com.caidr & 100,000+ & 4.08 \\
        10 & com.teckelmedical.mediktor & 50,000+ & 3.62 \\
    \hline
    
    \hline
        \multicolumn{2}{|l}{{\bf Average Statistics:}} &\multicolumn{2}{l|}{} \\
        
        \multicolumn{2}{|l}{ Avg. \# of Install}   &\multicolumn{2}{l|}{50,000+ }  \\ 
        \multicolumn{2}{|l}{ Avg. \# of Ratings}  &\multicolumn{2}{l|}{3.27} \\ 
  \hline
  \hline
\end{tabular}
\end{table}

\subsection{Analysis Methodology}
To have a wider perspective of the existing SymptomCheckers security, we perform comprehensive static and dynamic analyses to inspect apps source code and investigate the behaviour of the app during the runtime, respectively. We also conduct a user review analysis to determine the user perception of the analyzed SymptomCheckers.

\textbf{Source Code Analysis:} An APK is a mobile \emph{app package} file format supported by the Android operating system (OS) for distribution and installation. APK encloses all of the program's codes and it supports resources including .dex files, resources, assets, certificates, and manifest files, which are considered as the important objects in this SymptomCheckers static analysis. Since the APK is distributed in byte-code format, we conduct pre-processing by leveraging APKTools~\cite{apktool} to decompile the APK into Smali format and get all those files that will be useful in the following further analysis:

\begin{enumerate}
    \item {\bf Certificate Signing Mechanism.} The Android platform requires all APKs to be digitally signed with a certificate before it is uploaded to Google Play Store or installed on a device. Application signing simplifies developers to identify the app's author and to update their application without administering complicated permissions and interface. This process also becomes an insurance policy for developers in terms of apps integrity and the accountability of their apps' behavior. \cite{cert_sign2} illustrates the importance of this process to prevent adversaries from inserting malware into legitimate apps by modifying and repackaging apps on the apps market.
    
    Certificate signing in Android is conducted by integrating the identity and private key of the author into a digital certificate. The certificate is then attached to the app during packaging or before it is uploaded to the apps market. Since the certificate is publicly available during the distribution stage, the integrity and the accountability of the apps is highly dependent on the encryption and hashing mechanism tailored in this process.
    
   To evaluate the certificate signing mechanism adopted by the SymptomCheckers, we extract the {\tt CERT.RSA} file among all the files generated during the apps de-compilation. We customized a script leveraging Keytool\cite{keytool} to obtain encryption and hashing mechanisms as well as the length of the public key written into the certificate.
    
    \item {\bf Apps Permission Request.} The permission system is a core security architecture in the Android OS. All application requests to access sensitive data, system features, components or other sensitive resources in the operating system are managed by these systems. As an important point in security architecture, permission systems are very vulnerable to exploitation and become the attack surface of the Android system. Therefore, in this study, we included a permission system analysis to find potential dangerous permissions that could become a point of attack for malicious activities. For this purpose, we elaborated on {\tt Manifest.xml} on each app to find dangerous type permissions.
    
    \item {\bf Exported Component Analysis.} Android apps consist of several components: Activity, Service, Content Provider, and Broadcast Receiver. Activity analogous to the presentation layer and each activity represents each graphical user interface in the application while Service is a component that is responsible for handling background processes in the application. Content Provider and Broadcast receiver are responsible for exchanging data between apps and listening for system message requests called intents, respectively. These components collaborate to manage all functions in the apps. Commonly, a function in an app will be triggered by the user via the activity. The Android platform allows these components to be accessed and triggered from other applications by setting the exported status equal to True. However, this exported component is also a surface attack for malware to exploit an app. Melamed et al.,~\cite{exported_hack} demonstrates how these exported components can manipulate apps' components to compromise apps for malicious activities~\cite{exported_cwe}~\cite{zhao2019decade}. To identify the presence of an exported component in the SymptomCheckers, we use the Android Drozer\cite{drozer} to analyze the {\tt Manifest} file for each app. Drozer--a commonly used penetration testing tool--uses several checks to exploit vulnerabilities in mobile apps.
    
    \item {\bf Malware Detection}. To detect the presence of malicious codes in the SymptomCheckers, we scan APKs using {VirusTotal}~\cite{virustotal}.  {VirusTotal} is a multitude of malware scanning tools that provide a comprehensive result by aggregating more than 70 anti-virus engine and URL/domain blacklisting services.  The tools has been widely used to identify the emergence of malicious apps, executable files, application software as well as domains~\cite{ikram_vtscore}. To automate the scanning process, we take advantage of the API provided by {VirusTotal}, and create a script to upload all samples to the  {VirusTotal} repository.  

    Due to the file size restriction, among the total of 36 apps, we only manage to scan 25 (69.4\%) apps that have less than 32 MB in size through {\tt VirusTotal} API. We perform manual analysis through {\tt VirusTotal} interface for the rest of 11(30.5\%) apps including 8(22.2\%) apps that have a size larger than 32 MB and 4(11.1\%) apps that consist of multiple APK. Specifically for the apps with multiple APK, we conduct manual scanning to both the base and split APK to anticipate the malware that resides on the split APK. 
    
    \item {\bf Anti-Analysis Detection.} Anti-analysis technique refers to any means of evading, obscuring, or disrupting the analysis process by parties other than application developers. These techniques have both positive and negative sides. On the one hand, this technique is useful for hardening the apps and protecting the source code against analyzing and reproducing. On the other hand, this technique can be used by malware developers to evade basic analysis layers of application distribution services such as Google Play~\cite{apkid_ngock_tu}. Research in \cite{apkid_ren_he} found that 52\% of its malware samples leverage this technique to evade the analysis tools.
    
    To detect such behaviour in the SymptomCheckers, we use APKID~\cite{apkid_rednaga} to analyze the {\tt .dex} files obtained in decompiled APK. APKID returns at least one compiler name for each APK. If the apps leveraging any anti-analysis technique, the APKID will return several labels that we grouped as a manipulator, anti-virtual machine (vm), anti-debug, anti-disassembly and obfuscator.
    
    \item{\bf Tracker Detection.} The existence of third-party libraries and trackers on android apps has sparked concern about privacy violations for app users. These libraries can exchange information and infer user personal information based on demographic data and user behavior harvested during user interaction with the apps. Hence, in this study, we include the presence of these libraries as a factor associated with privacy violations of SymptomCheckers users.
    
    To reveal the existence of these libraries, we analyze the decompiled APK and comprehensively search sub-directories in decompiled APKs. These unique directories names correspond to the libraries embedded by apps’ developers in the source codes. We rely on our list of libraries to the previous research conducted in~\cite{hashmi2019longitudinal}~\cite{hashmi2019optimization}~\cite{AdBlocking} to filter and obtain the third-parties in SymptomChecking apps.

\end{enumerate}

\textbf{Runtime Behaviour Analysis:} We conduct runtime behaviour (also called dynamic analysis) to measure the security and privacy of SymptomCheckers apps during the runtime by capturing the traffic transmitted by Apps to the Internet. To avoid the Anti-Virtual Machine detection adopted by certain apps, we use a dedicated Android device and channelling the connection via MITMProxy~\cite{mitmproxy} to the WiFi access point. Since all the SymptomCheckers developed in SDK version 25 or above, we create a script to add self-signed security certificate exemption to read the traffic transmitted in HTTPS protocol. Although the security exemption was added to all apps, we failed to capture the traffic of 6 apps including 4 apps with multiple APK and 2 apps with a single APK.  All of these applications deny self-signed certificate exemption and required TLS handshake on the early phase of runtime, which causes the application to always be in a force stop condition.

Given that the number of SymptomCheckers is small, once the apps are installed on the device, we navigate the apps manually and observe all activities on the apps while MITMProxy intercepts the transmission between each app to the internet. The result of the interception is then saved in .dump form.

\textbf{User-review Analysis:}
A user review analysis was performed to capture the users' perceptions of SymptomCheckers. In this analysis, we have fetched Google Play store reviews (N = 76,817) of 30 apps. App reviews were categorized as positive (93.74\%) with rating = 3,4,5 and negative (6.26\%) with rating = 2 . We have only considered reviews of rating $\le2$ that were in total N = 4,807 negative reviews to understand the user's perception of app usability, mal-behaviour, privacy and security. To carry this analysis, we have used a dictionary of 60 keywords that mapped to 12 complaint categories (see Table~\ref{tab:user_perception}). For instance, keyword `{\tt force close}' mapped to `{\tt bugs}' under app's `{\tt usability}' category and `{\tt personal data}' mapped to `{\tt privacy}' for the `{\tt Privacy}' category. We found a total of N = 2,078 complaints across various categories that suggest the user's perception of each of them. \\

\section{\uppercase{Analysis}}
We explore the vulnerability of SymptomCheckers by identifying security gaps, suspicious behavior, level of communication security and user perspective. The results of the analysis present several aspects including; The strength of the encryption algorithm, exploitation of attack surface, intrusive permission, suspicious malware and anti-analysis appearance, adoption of secure communication protocol, and user-related security aspects.

\subsection{Certificate Signing Summary}
Based on the evaluation results of the certificate signing mechanism for the SymptomCheckers, we found 31 (86.1\%) apps signed using the SHA256 and RSA encryption mechanism. While the remaining 5 (13.8\%) apps were signed using a combination of SHA1 and RSA encryption mechanisms. During the exploration process, we received a warning from the key tool stating that apps with a SHA1 and RSA mechanism were categorized as having a fairly weak security level. SHA-1 was considered a weak signing mechanism after the first collision attack for the full SHA-1 white paper released by Google in \cite{sha-1_attack}.

\begin{table}
\centering
  \caption{Summary of Certificate Signing Analysis; Top: Certificate signing mechanism; Bottom: Public Key length written in Digital certificate.}
  \label{tab:certificate_sign}
  \begin{tabular}{|l|r|}
    \hline\hline
    
    {\bf Signature Algorithm}   &  {\bf \# of Apps, N=36 (\%)}  \\
    \hline
        SHA256 + RSA      & 31 (86.1\%)\\ 
        SHA1 + RSA (weak) & 5  (13.8\%)\\ 
  \hline
    \hline
    {\bf Key Length (bits)} &  {\bf \# of Apps, N=36 (\%))} \\
    \hline
        4096 & 15 (41.6\%)\\ 
        2048 & 21 (58.3\%)\\ 
  \hline\hline
\end{tabular}
\end{table}
We also include the length of the public key as part of the security measures in the SymptomCheckers certificate signing scheme. We are referring to the minimum standard public key length of 2048 bits published by \cite{key_length}. This document contains guidance on policy and security planning requirements for United States government agencies. 

As shown in Table \ref{tab:certificate_sign},  all SymptomCheckers have met minimum security standards of public key length, where 21 (58.3\%) apps adopting a public key with a length of 2048 bits and 15 (41.6\%) apps adopting 4096 bits. 

\subsection{Permission Analysis}
\begin{figure}[h]
    \centering
    \includegraphics[width=0.8\linewidth]{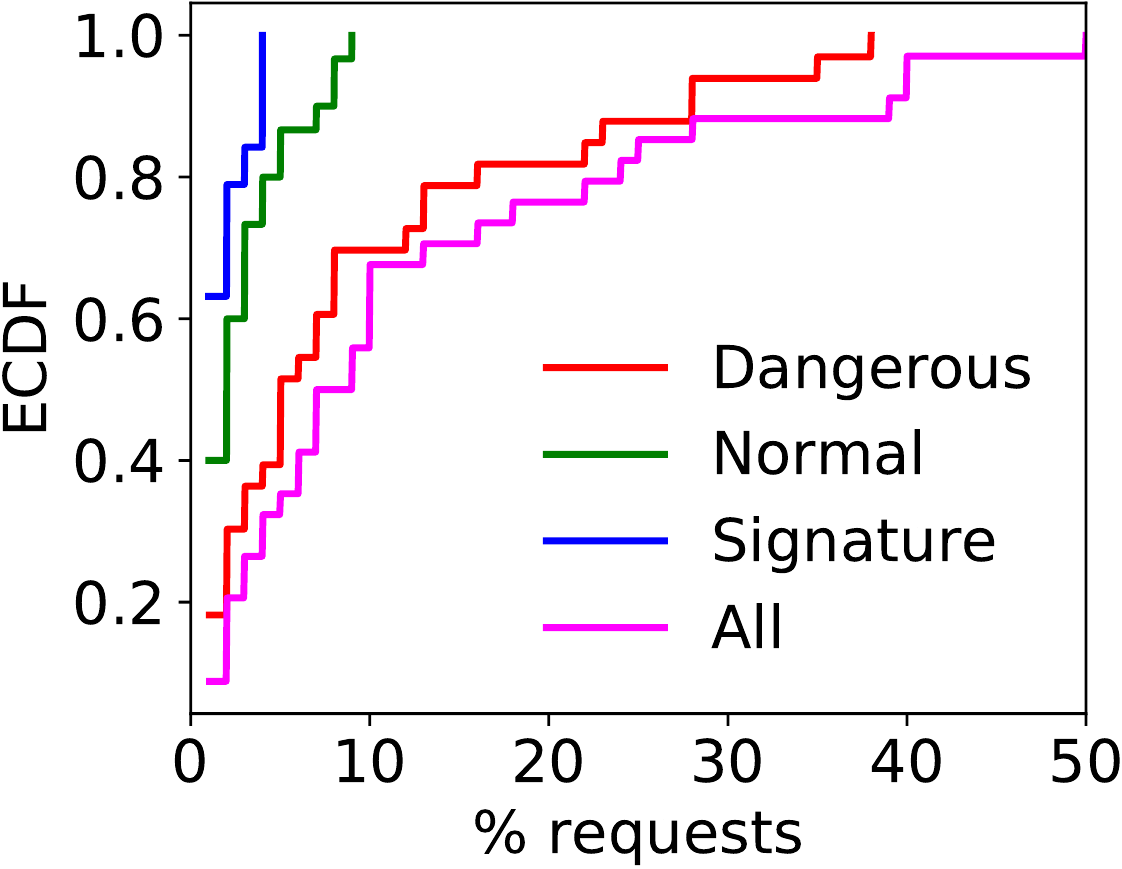}
    \caption{Cummulative Distributed Function of Permission request by SymptomCheckers. Dangerous permission approximated Total permission request.}
    \label{fig_ecdf_permission}
  \end{figure}

The dangerous permissions curve that approximate the total permissions curve in Figure \ref{fig_ecdf_permission} indicates that the number of dangerous permissions is close to the total permissions requested by the SymptomCheckers. Out of a total of 427 permission requests, 73 \% (312) permissions were categorized as dangerous, 20 \% (84) were categorized as normal, and 8 \% (33) permissions were categorized as signatures. The number of dangerous permissions is requested by 33 of the 36 apps on the SymptomCheckers list.

\begin{figure*}[h]
  \centering
  \includegraphics[width=1\linewidth]{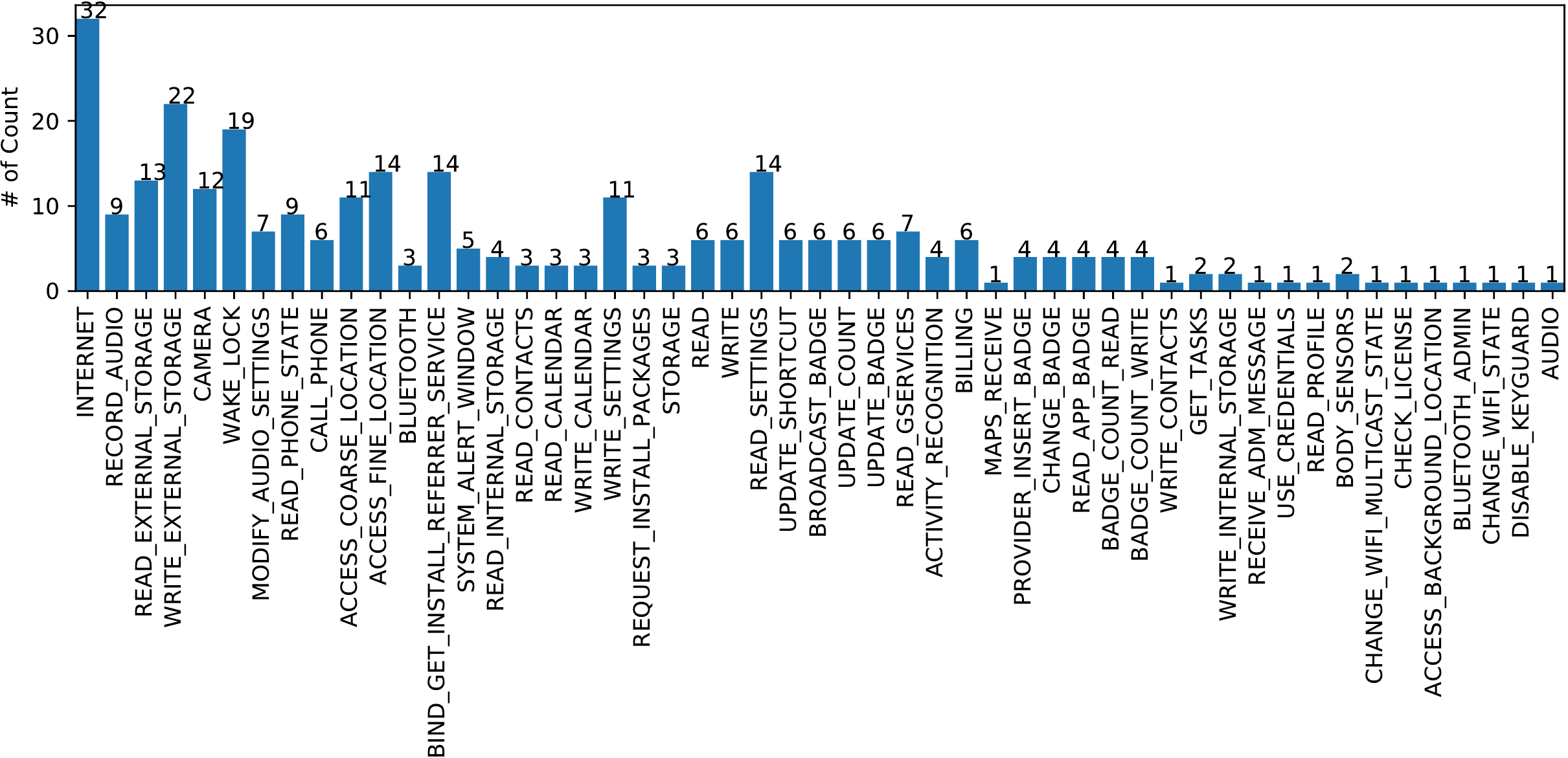}
  \caption{Overview of \emph{Dangerous} permissions requested by SymptomCheckers. {\tt INTERNET, WRITE\_EXTERNAL\_STORAGE, and WAKE\_LOCK} become the most popular permission request.}
    \label{fig:dangerous_permission}
\end{figure*}

From a total of 79 unique type permissions, 51 permissions in Figure \ref{fig:dangerous_permission}. {\tt INTERNET} is the most requested permission with 32 permissions, followed by {\tt WRITE\_EXTERNAL\_STORAGE} and {\tt WAKE\_LOCK} with 22 and 19 requests, respectively.

We then observe whether all the available list permissions are used by the functions in the respective SymptomCheckers. Therefore, we map the API calls or methods of each app with the permission requests in the Manifest using AXPLORER~\cite{axplorer}. As a result, all permission lists in the SymptomCheckers are requested at least once in the API calls or methods.

\subsection{Exported Component Analysis}
\begin{figure}[h]
    \centering
    \includegraphics[width=1\linewidth]{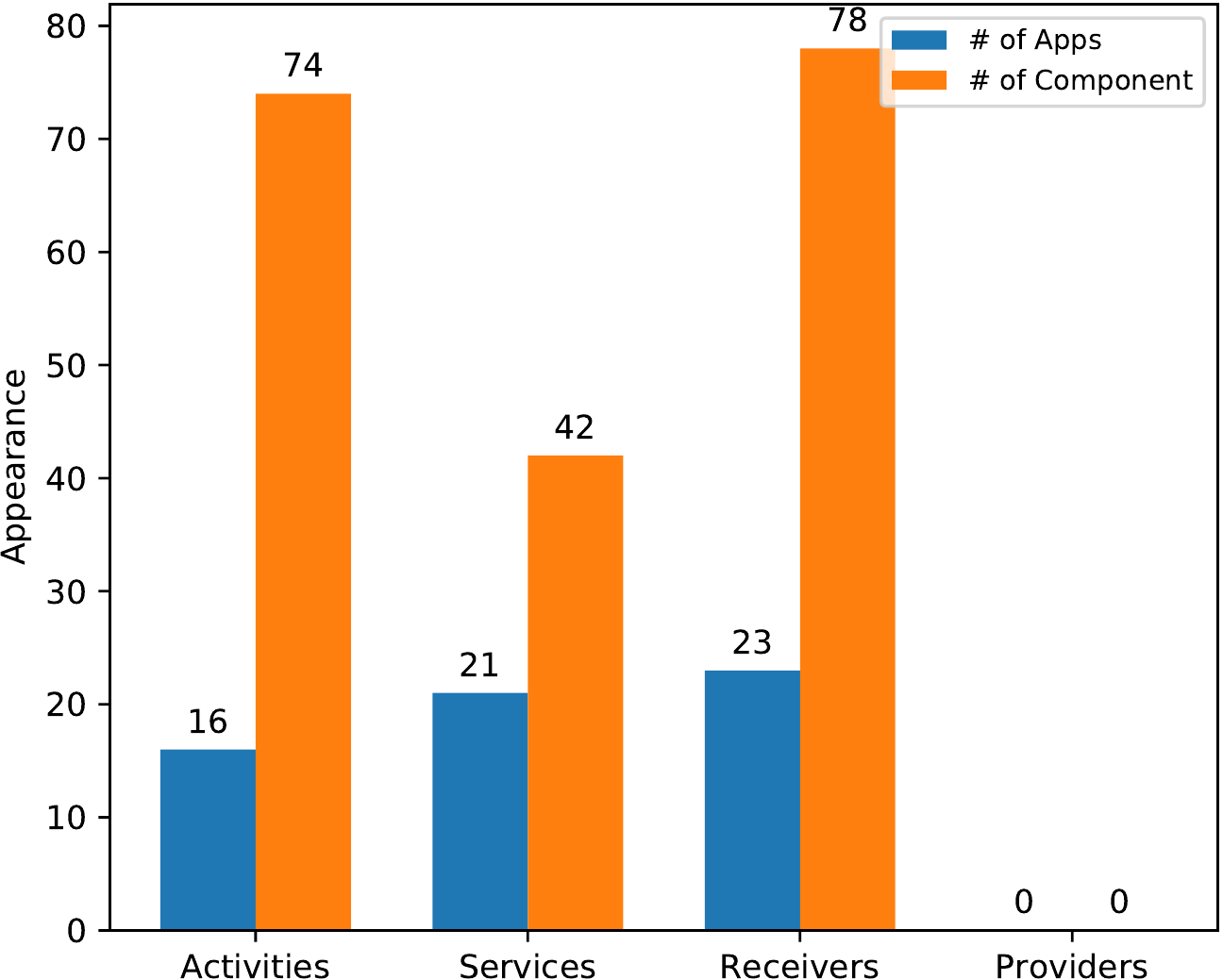}
    \caption{Results of SymptomCheckers' \emph{Exported Component} analysis. None of the apps containing Exported Content Provider.}
    \label{fig_exported_component}
  \end{figure}

 Figure \ref{fig_exported_component} depicts 44.4 \% (16) of the analyzed SymptomCheckers apps contain \emph{Exported Activities}. While, apps that have Exported Services and Exported Broadcast Receivers are 58.3 \% (21) and 63.86 \% (23) apps, respectively. None of SymptomCheckers apps containing Exported Content Providers. In total there are 74 Exported Activities, 42 Exported Services, 78 Exported Broadcast Receivers and none Exported Content Provider.

We then group the apps based on the number of Exported Components in each app as shown in Table \ref{tab:exported_component_sum}. In the table, it can be seen that 11 (30.5 \%) apps have Exported Component, 21 (58.3 \%) apps have Exported Services, and 18 (50 \%) apps have Exported Broadcast Receiver, in the range of 1 to 5 respectively. In addition, there are 9 (25 \%) apps that have Exported Components in the range of 6 to 10. We also notice massive Exported Activities by WebMD~\cite{WebMD} with 12 activities, indicating that there are 12 surface attacks that can be exploited by malicious activities in this app. 
For instance, WebMD~\cite{WebMD} consists of 66 activities, 10 Services, 11 Broadcast Receivers, and 6 Content Providers. This application is also listed as a SymptomCheckers with the highest number of installs (10,000,000+) with the number of reviews and ratings of 13,448 reviews and 64,152 users. 
\begin{table}
\centering
  \caption{Summary of \emph{exported components} analysis.}
  \label{tab:exported_component_sum}
  \small
  \begin{tabular}{|l|cc|cc|cc|cc|}
    \hline
    \hline
    {\bf Range} & \multicolumn{2}{c|}{\bf Activities } & \multicolumn{2}{c|}{\bf Services} & \multicolumn{2}{c|}{\bf Receivers} & \multicolumn{2}{c|}{\bf Providers} \\
    \cline{2-9} 
     & \# of Apps & \# of Comp & \# of Apps & \# of Comp & \# of Apps & \# of Comp & \# of Apps & \# of Comp \\ 
    \hline
        
        $>$10 & 1 &  12 &  0 &  0 &  0 &  0 &  0 &  0\\ 
        6-10 & 4 &  36 &  0 &  0 &  5 &  30 &  0 &  0\\ 
        1-5 & 11 &  26 &  21 &  42 &  18 &  46 &  0 &  0\\ 
        0 & 20 &  0 &  15 &  0 &  13 &  2 &  36 &  0\\ 
    \hline\hline
\end{tabular}
\end{table}

\subsection{Malware Presence}
To simplify the interpretation of {\tt VirusTotal} result, we introduce VTScore which represents the number of detection by anti-virus engine in the {\tt VirusTotal}. The score range is between 0 to 70 where the maximum score is equal to the number of anti-virus engines in {\tt VirusTotal}.

Upon scanning SymptomCheckers using {\tt VirusTotal}, we find 4 (10.81\%) apps detected to contain malware with a VTScore of 1. Those are the apps that manually scanned through the {\tt VirusTotal} interface including  MiA SymptomChecker~\cite{miaSymptomCheckr2019}, 
Mediktor~\cite{Mediktor}, HealthNav~\cite{healthnav}, and Mayo Clinic~\cite{mayoclinic}.
Those are popular apps with a number of installs above 10,000+ and has been released for more than three years except for MiA SymptomChecker  
which was only installed by 100+ users and was only released in December 2019. 50\% of the apps have a rating score above the average of the entire sample except for the  
MiA SymptomChecker and Mediktor which have a rating score of 0 and 3.62 respectively.

Further identification finds that all the VTScore were contributed by {\tt Sangfor Engine Zero} anti-virus.{\tt Sangfor Engine Zero} (SAVE Engine) is Artificial Intelligence based anti-virus engine developed by Sangfor Security Team. This engine leverages machine learning technology to analyze and synthesize the input data. This engine has become a part of {\tt VirusTotal} since November 2019. Unfortunately, {\tt Sangfor Engine Zero} detection results only provide a general label of malware so we cannot identify the malware type and family.

Based on the previous researches in~\cite{DREBIN} and ~\cite{ikram_vtscore} that establish minimum confidence VTScore of 2 and 5 respectively, and the reputation of {\tt Sangfor Engine Zero} explored in \cite{zangfor_misc1,zangfor_misc2}, we consider that the detection result has a possibility of being false-positive.


\subsection{Anti-Analysis Techniques}

\begin{figure}[h]
    \centering
    \includegraphics[width=\linewidth]{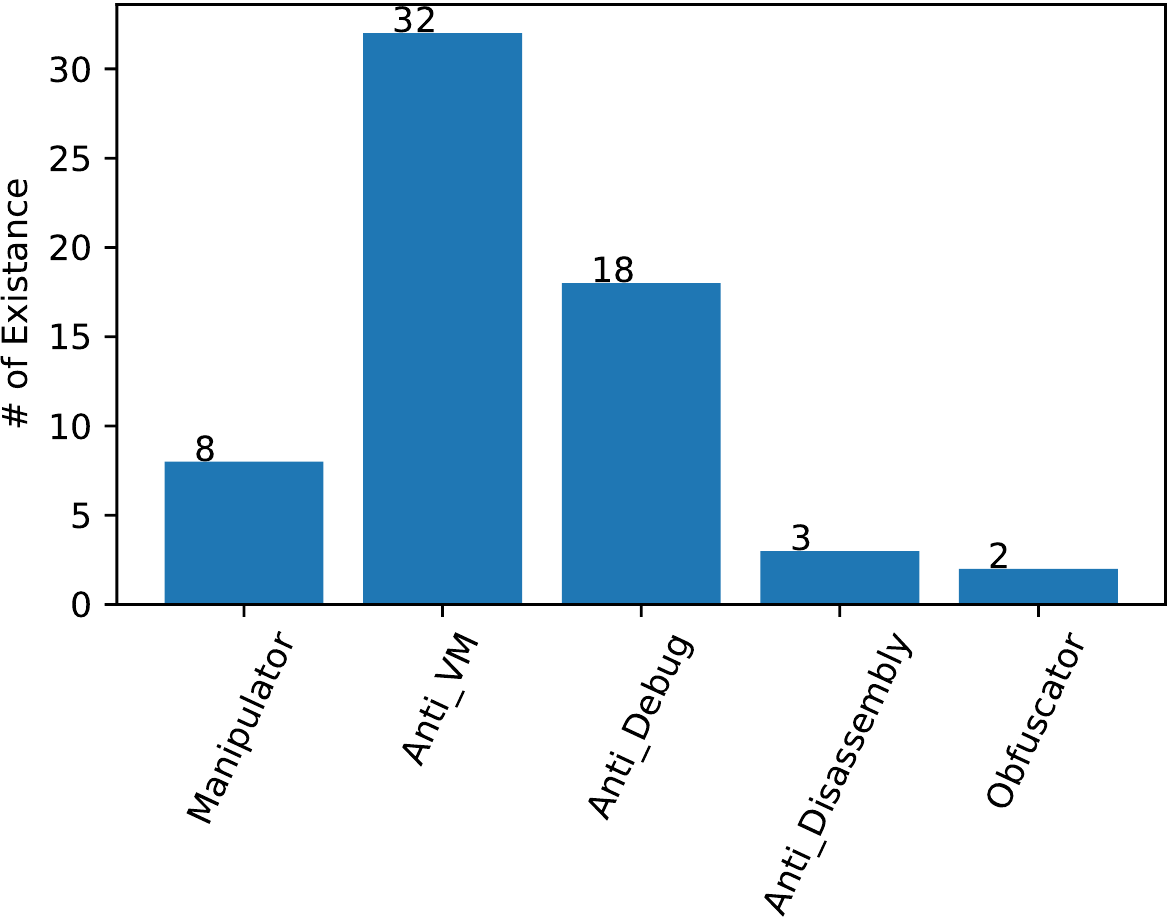}
    \caption{Anti-Analysis techniques adopted by SymptomCheckers. 89\% SymptomCheckers apps leveraging anti-VM technique to detect emulated environment. }
    \label{fig_apkid_result}
  \end{figure}

The Evaluation results of the SymptomCheckers show that 89\% of SymtomCheckers have adopted at least one anti-analysis techniques in their source codes. The details of the measurement results are described as follows: 
\begin{itemize}

    \item \textbf{Manipulator}: Each APK consists of {\it Dalvix Executable (.dex) } files converted from {\it Java.class} using (originally) {\it dx} compiler. We marked the apps containing the manipulator if the {\it .dex} files were created using other than {\it dx} compiler. There are two condition that make the manipulator label can be given; first, when  the original {\it .dex} files created using {\it dx} compiler is modified using modificator library such as {\it dexmerge}, and second, when {\it .dex} files created from reverse-engineered source code using {\it dexlib} library that commonly used by decompiler tools such as apktool or smali~\cite{apkid_manipulator1}.  Besides being used as a logical reference for the {\it Dalvik Virtual Machine (DVM)} or {\it Android Runtime (ART)}, {\it .dex} files also store a history of changes experienced by the apps. APKID checked that history by analyzing {\it Map Ordering Type} of the {\it .dex} files since the code sequence resulted from original {\it .dex} compiler, {\it dexmerge, dexlib or dex2lib} is different~\cite{apkid_manipulator3}. 
    
    As shown in figure \ref{fig_apkid_result}, we found that 8 (22\%) SymptomCheckers leverage {\it dexmerge} compiler to protect their source codes.
    
    
    \item \textbf{Anti Virtual Machine}:  This technique is used to detect whether the apps are running on the emulator or real devices. The emulator detection aims to increase the difficulty level of apps running on emulators which impedes certain reverse-engineering tools and techniques. The reverse engineer required additional tools to deceive the emulator detection or utilized the physical devices to reverse the source code. There are several indicators to check whether the device is emulated~\cite{apkid_anti-debug}. The first indicator can be found in file {\it build.prop} that contains a static list of Build API methods including {\it Build.Fingerprint, Build.Hardware, Build.Device} and other devices properties. The other set static indicator can be checked from {\it Telephony manager} which have fixed API values of Android emulators including {\it getNetworkType(), getNetworkOperator(), getPhoneType() } and other network-related properties.
    
    We found 32 (89\%) apps use anti-virtual machine techniques to check the emulator presence based on various indicators in {\it build.prop} and {\it Telephony Manager}.

    \item \textbf{Anti Debug}: There are two types of anti-debuging which are preventive and reactive. Preventive anti-debuging prevents any kind of debugger activities from the first time. While reactive anti-debuging started with detecting debuggers and reacting in certain ways, such as triggering hidden behavior or terminating the apps. There are two levels of debugging as well as anti-debugging protocol in Android~\cite{apkid_anti-debug}. First, the debugging can be conducted in Java level using Java Debug Wire Protocol (JDWP) which is used as communication protocol between debugger and Java Virtual Machine. In this level we can trace anti-debugging setup by checking the {\it  debuggable flag} in {\it ApplicationInfo} or by checking the {\it timer checks routine}. Second, we can conduct debugging in native layer level by using {\it ptrace} in Linux {\it system call}. This anti-debugging level check is also known as a traditional anti-debugging process.
    
    We found 18 (50\%) SymptomCheckers tailor anti-debug techniques to avoid analysis tools. All of those apps detect debuggable state by activated {\it isDebuggerConnected()} routine in {\it android.os.Debug class}, which is part of JDWP anti-debugging level technique.

    \item \textbf{Anti Disassembly}: Anti-disassembly could be the most advanced mechanism in the anti-reversing domain. The developer needs to craft their code in order to protect their byte-code from being disassembled by the reverse engineer. A common technique to protect the byte-code in Android is to create the important code segment in C or C++ using Native Development Kit (NDK)\cite{apkid_diss1}. NDK provides platform libraries to manage native activities and access physical device components\cite{apkid_diss2_ndk}. NDK uses {\it CMake} as a native library compiler that creates a different {\it byte-code} structure compared to the code written in Java or Kotlin. Hence, it impedes common Android tools such as Apktool or Smali to disassemble the {\it byte-code}. It required advanced reverse engineering that was familiar to ARM processor architecture, Assembler language, Java Native Interface (JNI) convention and Application Binary Interface (ABI) compiler to decompile the {\it byte-code}. More advanced technique used by malware developers to evade disassembly tools explained by ~\cite{apkid_diss3_adv}. The technique leveraging {\it jmp} and {\it call} command in {\it byte-code} level to direct the instruction flow to a certain location with a constant value, or directing the flow to the same target memory location. This technique will produce a false listing of source code when it disassembles using a decompiler tool.  
    
    We found 3 (8\%) SymptomCheckers use anti-disassembly techniques, including AITibot~\cite{aitibot}, Appstronout~\cite{appstrounout} and Mayo Clinic ~\cite{mayoclinic}. Analysis of those apps returns the value of ``illegal class name'', indicating the decompiler result violating the standard structure of Java or Kotlin.

    \item \textbf{Obfuscator}: Code obfuscation is the process of modifying code into meaningless phrases by renaming or encrypting the file names, methods, or strings without reducing the code functionality. The aim is to protect the executable file from being analyzed or being reversed by an unauthorized party. Other than renaming and encrypting, ~\cite{apkid_obs1} describes more advanced techniques to conduct code obfuscation including control flow obfuscation and instruction pattern transformation. Proguard~\cite{apkid_obs2_proguard} and Dexguard~\cite{apkid_obs3_dexguard} are two most common tools utilized in Android programming obfuscation. Proguard is a free and open-source tools that provide an optimizer for general Java byte-code. It enables developers to obfuscate, compress, and optimize desktop, embedded and mobile Android applications. Dexguard is proprietary tool designed to protect Android applications. It provides multi-layer protection against the static and dynamic analysis of byte-code, manifest, and all other resources included in distribution packages.
    
    We found 2 (6\%) SymptomCheckers deploying obfuscators. Analysis results on Healhtily~\cite{mdyour_healtyly} return 'unreadable field names' and 'unreadable method names' indicating that a certain number of field names and method names in that app were renamed or encrypted. However, APKID failed to identify the tools used to conduct obfuscation. While analysis on Mayo Clinic~\cite{mayoclinic} identifies the apps adopting Dexguard to conduct the obfuscation.

\end{itemize}

\subsection{Third-party Ads and Tracking Analysis}
\label{third_party}
\begin{figure*}[h]
  \centering
  \includegraphics[width=1\linewidth]{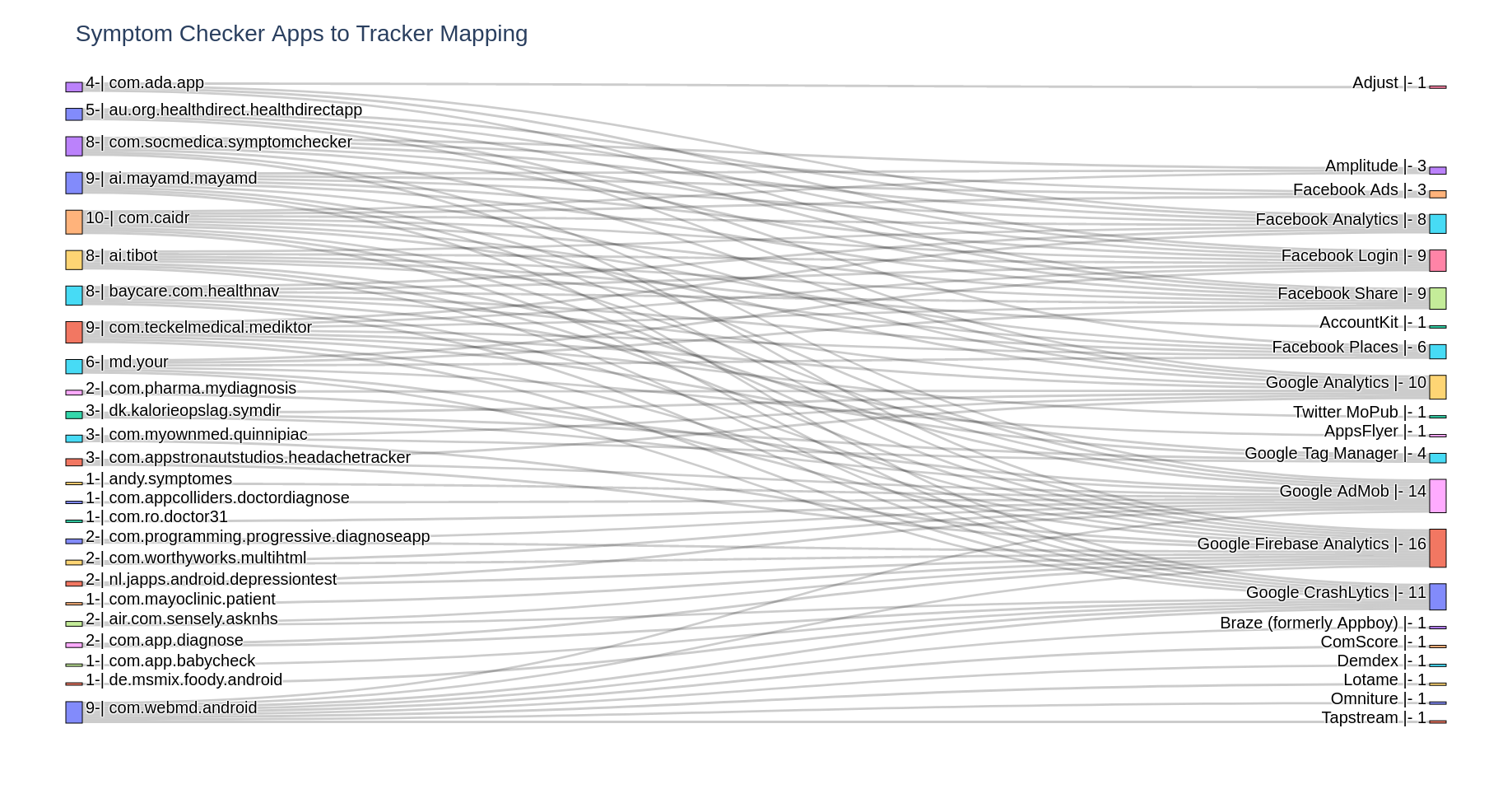}
  \caption{Apps to third-party libraries mapping. Left column indicating the number of third-party libraries adopted by SymptomChekcers, while right column figuring the name of third-party library and the number of usage by SymptomCheckers.}
    \label{fig_tracker}
\end{figure*}

The results of the tracking library analysis show that 69.4\% (25) SymptomCheckers adopt at least one tracker. There are even 22.2\% (8) apps leveraging more than 5 tracking libraries in their code. The application with the highest number of tracking libraries was listed by {\tt com.caidr.apk} with 10 trackers. {\tt com.caidr} is a popular app with over 100,000+ installations and rating of 4.1 . These apps consist of 26 activities where almost 50\% of the activities are proposed to handle third-party libraries including tracker. More details about the results of this measurement can be seen in Figure \ref{fig_tracker}.

The detection results show that the adoption of a third-party library on SymptomCheckers is dominated by two large company groups, namely Google which includes Google Analytics, Google Tag Manager, Google Admob, Google Firebase Analytic, and Google CraschLytic, and the Facebook group consisting of Facebook Ads, Facebook. Analytic, Facebook Login, Facebook Share, and Facebook Place. Totally, Google groups are dominated by 55 libraries attached to the SymptomCheckers, followed by Facebook groups with 35 libraries. In this research, Google Firebase Analytic became the most popular third-party library that is adopted by 44.4\% (16) of SymptomCheckers.

Firebase Analytics is a library used to infer personal user information including gender, age range, and location, based on demographic and behavior of users when interacting with applications\cite{UserPropertiesAnalytics,FireBaseUSerProperties}. This library is integrated with Google Analytics which on the surface can be detected only harvesting network data and device information\cite{PaidAppsTracking}. Although each library collects limited information, when all these libraries exchange information, it raises concerns about privacy violations because these third-party libraries can infer personal information based on the behavior and demographic data exchanges.

\subsection{Traffic Analysis}
Upon the interception process, we then convert the dump file into HTTP/S Archive (.har) file to simplify traffic analysis of each SymptomCheckers. We direct our analysis of security by extracting the secure communication line and privacy factors by observing the potential of privacy leaks.

\begin{figure*}[th!]
\centering
    \subfloat[HTTP and HTTPS Traffic]
    {\includegraphics[width=0.9\columnwidth]{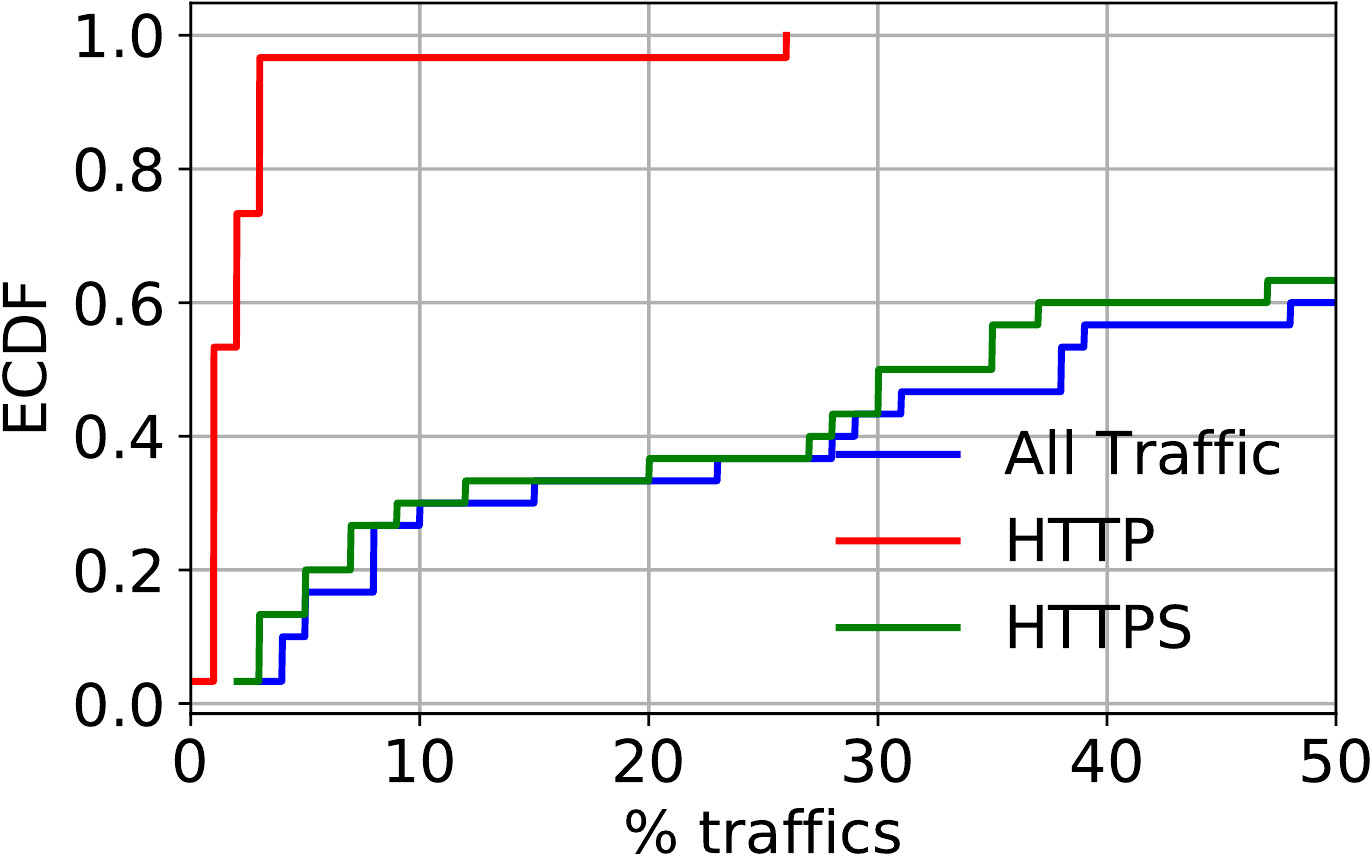}\label{fig:http_https}}\quad
    \subfloat[First Party and third-party Traffic]
    {\includegraphics[width=0.9\columnwidth]{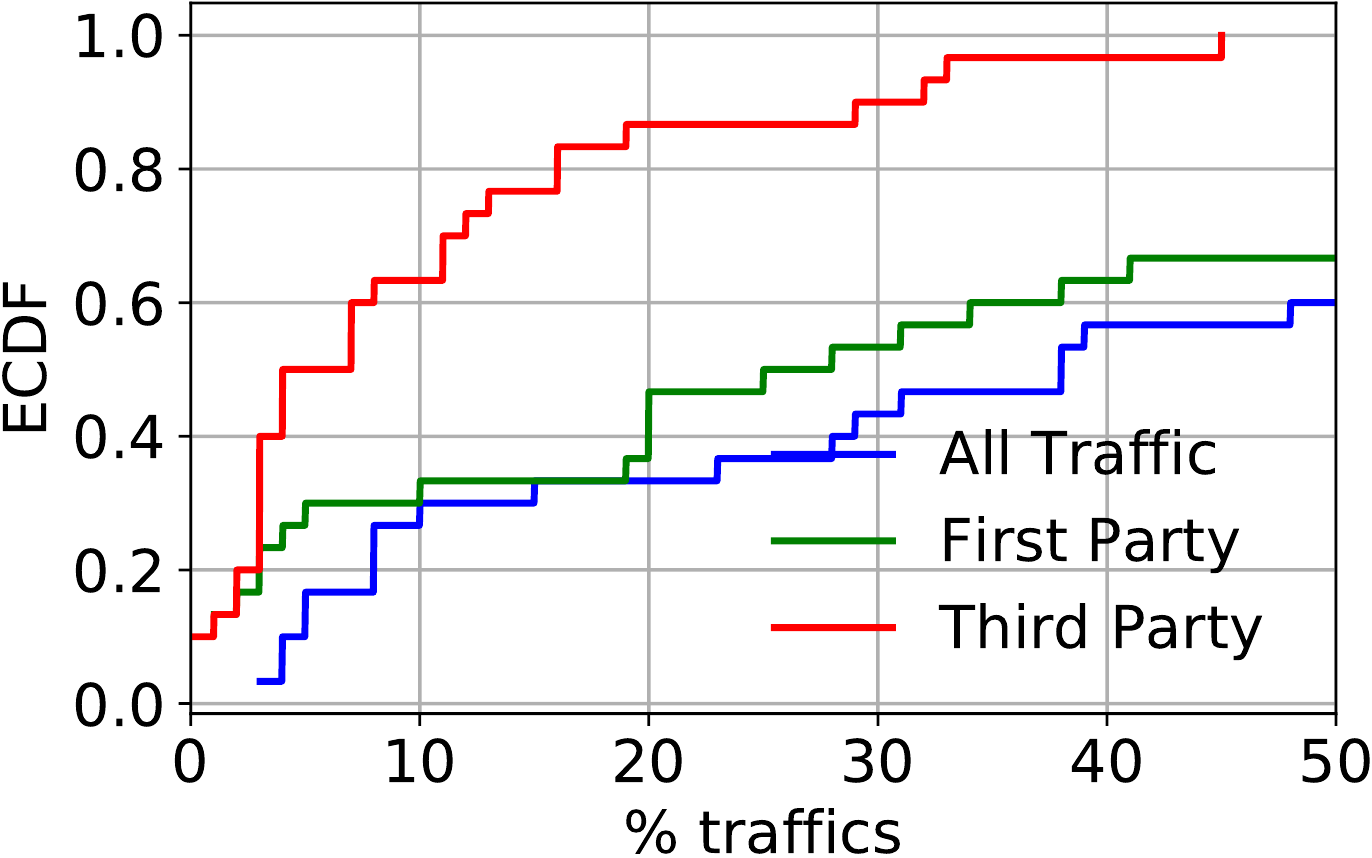}\label{fig:first_third}}
    \caption{Cumulative distribution function of Traffic Captured in SymptomCheckers. 95\% SymptomCheckers adopting encrypted HTTPS protocol and 81\% of the traffic are directed to the first-party or supporting domains.}
	\label{fig:traffic_analysis}
\end{figure*}

From the security factor, we analyzed the percentage of the encrypted HTTPS protocol adoption on the SymptomCheckers' communication lines. For this purpose, we extract intercepted unique URLs and identify the type of transmission protocol used. Based on the Cumulative Distribution Function (CDF) in Figure \ref{fig:http_https}, it can be seen that the HTTPS protocol adoption curve approaches the total communication line curve. This shows that most of the communication traffic for SymptomCheckers have adopted the HTTPS protocol. However, from a total of 1584 unique URLs detected, 74 (5\%) communications were passed through the HTTP protocol. 3 SymptomCheckers communicate 50\% of their traffic un-encrypted via HTTP. 

To observe potential privacy leaks, we consider the existence of third-party libraries adopted by each SymptomCheckers. Hence, we refer to Section \ref{third_party} to create a list of third-party libraries and compare it to the intercepted unique URLs. We then categorize the corresponding URLs as traffics that are directed to third-party domains and the rest is traffics that are directed to domains provided by the app developer (first-party) or domains that are used to support apps functionality. The percentage of traffic leading to a third-party domain is an indicator of a possible privacy leak.

The first-party traffic curve approximates the total transmission traffic curve in Figure \ref{fig:first_third} indicating that most of the traffic in the SymptomCheckers is directed to the first-party domain or apps supporting domain. However, out of 1,584 total traffic intercepted, there are 301 (19 \%) unique URLs that point to a third-party domain. Our further observation found that there were 8 (22\%) apps that had traffic leading to a third-party with more than 50\% of the total traffic. This indicates that the app uses more than 50\% of its functionality for third-party libraries.

\begin{table}
\centering
  \caption{Summary of Traffic Analysis. 19\% traffic directed to third-party domains and 1\% of the traffic relying on un-encrypted HTTP protocol.}
  \label{tab:trafic_analysis}
  \small
  \tabcolsep=0.08cm
  \begin{tabular}{|l|r|r|r|}
    \hline
    \hline
    & {\bf Unique URL} & {\bf HTTP} & {\bf HTTPS}\\ 
    \hline
        {\bf First-party} & 1,283 (81\%) & 69(4\%) & 1,214 (77\%)\\
        {\bf Third-party} & 301 (19\%) & 5 (1\%) & 296 (19\%)\\
        \hline
        {\bf Total} & 1,584 (100\%) & 74 (5\%) & 1,510 (95\%)\\
    \hline\hline
    \end{tabular}
\end{table}
Table \ref{tab:trafic_analysis} summarises our traffic analysis. 
Of 1,584 intercepted unique URLs, there are 1,510 (95\%) traffic transmitted over the encrypted HTTPS protocol line and 74 (5\%) traffic transmitted via the un-encrypted HTTP protocol. In addition, of the total Unique URLs, there are 1,283 (81\%) URLs pointing to first-party or supporting domains and 301 (19\%) URLs pointing to third-party domains. In more detail, 69 (4\%) of traffic to the first-party domain is transmitted via the un-encrypted HTTP and 1,214 (77\%) of traffic to the first-party domain is transmitted via the encrypted HTTP. While, 5 (1\%) and 296 (19\%) unique URLs were directed to the third-party domain via HTTP and HTTPS, respectively.

\begin{table*}[!ht]
  \caption{Analysis of users' perception of SymptomCheckers analyzed through user reviews categories and keywords. }
  \label{tab:user_perception}
  \centering
  \tabcolsep=0.08cm
  \begin{tabular}{|lrr | p{5.5cm}|}
    \hline
    \hline
    {\bf Complaint Category}   &  {\bf \#Comp (\%)} & {\bf \# Apps (\%)}& \textbf{Case-insensitive, Searched Keywords} \\
    \hline
     {\bf Usability} &\multicolumn{3}{c|}{} \\
    \hline
        Bugs  & 429(20.64)  & 12(41)& force close; crash; bug; freeze; glitch; froze; stuck; stick; error; disconnect; not work; not working \\ 
        Battery  & 74(3.56)  & 9(31) & battery; cpu; processor; processing;  ram ; memory\\
        Mobile Data & 87(4.19) & 10(34)& mobile data; gb; mb; background data; \\ 
  \hline
  {\bf Mal-behaviour} &\multicolumn{3}{c|}{} \\
  \hline
        Scam & 28(1.35) & 6(21)& scam; credit card; bad business; bad app\\
        Adult & 23(1.11) & 6(21) & porn; adult; adult ad\\
        Offensive/Hate & 2(0.10) & 2(7) & hate; offensive; sexist; LGBT; trolling; racism; offensive; islamophobia; vile word; minorities; hate speech; shit storm\\ 
   \hline
   {\bf Privacy} &\multicolumn{3}{c|}{}\\
  \hline
        Privacy & 53(2.55) & 7(24) & privacy; private; personal details; personal info; personal data\\
        Ads & 1280(61.60) & 24(83) & ads; ad; advertisement; advertising; intrusive; annoying ad; popup; inappropriate; video ads; in-app ads\\
        Trackers & 43(2.07) & 10(34) & tracker; track; tracking\\ 
  \hline
  {\bf Security} &\multicolumn{3}{c|}{}\\
  \hline
        Security & 12(0.58) & 4(14)& security; tls; certificate; attack\\
        Malware & 5(0.24) & 3(10) & malware; trojan; adware; phishing; suspicious; malicious; spyware \\
        Intrusive Permissions & 42(2.02) & 6(21) & permission\\
  \hline\hline
\end{tabular}
\end{table*}
\subsection{User Perception Analysis}
Users often complained about apps' usability (see Table~\ref{tab:user_perception}) where bugs, battery and mobile data issues caused frustration. For instance, 429(20.64\%) complaints were made for 12 (41\%) apps where users raised their concerns and requested to fix the usability issues to improve the user experience. While some might find the app useful but an apparent usability issue may hinder useful experience. Some users stated that "great app is very useful; however I have gotten a lot of 'force close' messages please fix.", "would not load pictures and kept freezing" or "had to uninstall as they kept crashing every time you open the app". 

Apps mal-behaviour had captured minor users' attention where such apps were listed as scams and raised concerns about presenting adult or offensive content. User stated as "No way to verify that this is a legitimate app. No clear connection to the NHS. Looks like a scam to steal personal information". 

Notably, under the privacy category users mainly complained about in-app advertisements 1280 (61.60\%) for 24 (83\%) reviewed apps. Users elaborated "don't want seeing this app advertisement through my phone", "I need to answer your questions and all I see is advertisement..." or "... I cannot find anything not my blood type not my last appointment everything is advertisement...". Comparatively user reviews suggest little understanding of apps privacy and tracking behaviour. However, some users stated "privacy policy is bad", "sharing personal data is not cool" or "removed the app after finding out that it shared private health data with third-parties". 

Similar to privacy users' perception of apps security was also limited. There were minor complaints made for app security, presence of malware and app requiring intrusive permissions as shown in Table~\ref{tab:user_perception}. Users mentioned their concerns for specific apps as "security risk, asks for personal details that would facilitate identity theft. if in doubt be cautious." or "spyware, why does it need my device serial number; my location; read SD card; to record audio when I get up and go to sleep; all combined with full internet access?". 

We found that 20.6\% and 61.6\% of users' complaints were related to usability issues such as bugs (20.64\%) and advertisements, respectively. A fraction of users complaints approx. 3\% (57) are related to security issues in 13 SymptomCheckers. 

\section{Related Work}
	Our study builds on previous work in static and dynamic analysis of mobile apps
	Static analysis is \textit{parsing the source-code of mobile apps}, while dynamic analysis is studying the behavior of apps using runtime network analysis~\cite{ikram_vtscore}. 

	Static analysis can be performed by either static data flow analysis \cite{AdBlocking}, ~\cite{PScout}, \cite{IkramBK19}, or symbolic execution \cite{SymDriod}. Previous studies have highlighted several PII leaks \cite{tangari2021mobile} \cite{DriodJust} albeit with higher false positives. The cases where Android apps load the code at run time only (as presented by Lindorfer et al. \cite{lindorfer2014andrubis}), can result in higher false negatives, while, symbolic execution can be too time-intensive to be practical \cite{ren2018appversions}.
	Au et al. \cite{PScout} and Leontiadis et al. \cite{Leontiadis2012} leverage static analysis to inspect app permissions and their associated system calls. By developing a version-independent static analysis tool called PScount, Au et al. \cite{PScout} found that around 22\% of the non-system permissions are redundant. While Leontiadis et al. \cite{Leontiadis2012} proposed a feedback control loop framework, that can alter the level of privacy protection on mobile phones by examining 250 thousand applications. 
	

Dynamic analysis suffers from false overhead, adding to its challenges when running on a large scale. Moreover, it accumulates run-time overhead, adding to its challenges when running on a large scale.
%
	%
	Perhaps, a much closer work to our's is by Ren et al.~\cite{ren2018appversions}, where they monitored the network traffic to detect personally identifiable information (PII) leaks of 512 Android apps across different versions. They show that apps leak PII to more third-party domains over time. We build on this work by conducting a compliance analysis of apps across different versions by taking the union of leaks that are detected in static and dynamic analysis.

\section{\uppercase{Conclusion}}

The increasing number of mobile SymptomCheckers available on
apps’ markets such as Google Play and the growing number
of complaints raised by users indicate serious security and privacy as well as
 usability issues thus necessitate the urge to analyse this
unexplored eco-system. The average mobile user rates SymptomCheckers positively even when 10\% have malware presence. According to our study, 3\% of negative reviews are related
to (or concerned with) the security and privacy issues of SymptomCheckers. Our app review analysis suggests an alarming situation as users show minimal interest or awareness of SymptomCheckers privacy and security. 
The results found in this paper reveal serious security and privacy issues in Google Play. 
We believe that our work could be extended to study the qualitative analysis such as users satisfaction and effectiveness of SymptomCheckers' functionalities.

\bibliographystyle{splncs04}
\bibliography{symptom_checker}

\begin{thebibliography}{10}
\providecommand{\url}[1]{\texttt{#1}}
\providecommand{\urlprefix}{URL }
\providecommand{\doi}[1]{https://doi.org/#1}

\bibitem{aitibot}
AITibot: Understand skin problems - ai tool for skin.
  \url{https://play.google.com/store/apps/details?id=ai.tibot} (2021)

\bibitem{apktool}
Apktool: A tool for reverse engineering 3rd party, closed, binary android apps.
  \url{https://ibotpeaches.github.io/Apktool/} (2020)

\bibitem{appstrounout}
Appstrounout: Headache tracker - migraine \& headache log.
  \url{https://play.google.com/store/apps/details?id=com.appstronautstudios.headachetracker}
  (2021)

\bibitem{DREBIN}
Arp, D., Spreitzenbarth, M., Hübner, M., Gascon, H., Rieck, K.: Drebin:
  Effective and explainable detection of android malware in your pocket. In:
  NDSS (2014)

\bibitem{PScout}
Au, K.W.Y., Zhou, Y.F., Huang, Z., Lie, D.: Pscout: Analyzing the android
  permission specification. In: CCS (2012)

\bibitem{axplorer}
Backes, M., Bugiel, S., Derr, E., McDaniel, P., Octeau, D., Weisgerber, S.: On
  demystifying the android application framework: Re-visiting android
  permission specification analysis. In: USENIX Sec (2016)

\bibitem{apkid_ngock_tu}
{Chau}, N., {Jung}, S.: An entropy-based solution for identifying android
  packers. IEEE Access  (2019)

\bibitem{miaSymptomCheckr2019}
Checker, M.S.: Mia symptom checker.
  \url{https://play.google.com/store/apps/details?id=
  com.nextgenblue.symptomchecker} (2021)

\bibitem{DriodJust}
Chen, X., Zhu, S.: Droidjust: Automated functionality-aware privacy leakage
  analysis for android applications. In: WiSec (2015)

\bibitem{mayoclinic}
Clinic, M.: Mayo clinic medical.
  \url{https://play.google.com/store/apps/details?id=com.mayoclinic.patient}
  (2021)

\bibitem{zangfor_misc1}
Community, S.: Incorrect detection at virus total by sangfor engine zero
  antivirus.
  \url{https://community.sangfor.com/forum.php?mod=viewthread\&tid=2854} (2020)

\bibitem{key_length}
CSRC: Recommendation for key management, part 1: General (revision 3), computer
  security resource center.
  \url{https://csrc.nist.gov/publications/detail/sp/800-57-part-1/rev-3/archive/2012-07-10}
  (2016)

\bibitem{exported_cwe}
CWE-926: Cwe-926: Improper export of android application components.
  \url{https://cwe.mitre.org/data/definitions/926.html} (2020), accessed:
  18/12/2020

\bibitem{apkid_diss2_ndk}
Developer, A.: The android ndk: toolset that lets you implement parts of your
  app in native code, using languages such as c and c++.
  {https://developer.android.com/ndk} (2020)

\bibitem{apkid_obs2_proguard}
Developer, A.S.: Shrink, obfuscate, and optimize your app.
  \url{https://developer.android.com/studio/build/shrink-code} (2020)

\bibitem{drozer}
F-Secure-Labs: Drozer: Comprehensive security and attack framework for android.
  \url{https://labs.f-secure.com/tools/drozer/} (2020), accessed: 18/12/2020

\bibitem{apkid_manipulator3}
Fenton, C.: Building with and detecting android's jack compiler.
  \url{https://calebfenton.github.io/2016/12/01/building-with-and-detecting-jack/}
  (2016)

\bibitem{UserPropertiesAnalytics}
Google-Analytic: User properties | analytics.
  \url{https://developers.google.com/analytics/devguides/collection/firebase/android/properties}
  (2020)

\bibitem{FireBaseUSerProperties}
Google-Firebase: Predefined user dimensions - firebase.
  \url{https://support.google.com/firebase/answer/6317486} (2016)

\bibitem{apkid_obs3_dexguard}
Guardsquare-Mobile-Application-Protection: Dexguard: Android app security -
  protecting android applications and sdks against reverse engineering and
  hacking. \url{https://www.guardsquare.com/en/products/dexguard} (2020)

\bibitem{hashmi2019longitudinal}
Hashmi, S.S., Ikram, M., Kaafar, M.A.: A longitudinal analysis of online
  ad-blocking blacklists. In: 2019 IEEE 44th LCN Symposium on Emerging Topics
  in Networking (LCN Symposium). pp. 158--165. IEEE (2019)

\bibitem{hashmi2019optimization}
Hashmi, S.S., Ikram, M., Smith, S.: On optimization of ad-blocking lists for
  mobile devices. In: Proceedings of the 16th EAI International Conference on
  Mobile and Ubiquitous Systems: Computing, Networking and Services. pp.
  220--227 (2019)

\bibitem{apkid_ren_he}
He, R., Wang, H., Xia, P., Wang, L., Li, Y., Wu, L., Zhou, Y., Luo, X., Guo,
  Y., Xu, G.: Beyond the virus: A first look at coronavirus-themed mobile
  malware (2020)

\bibitem{Ada}
Health, A.M.: Ada – check your health.
  \url{https://play.google.com/store/apps/details?id=com.ada.app} (2021)

\bibitem{mdyour_healtyly}
Healthily: Self-care \& health journal.
  \url{https://play.google.com/store/apps/details?id=md.your} (2021)

\bibitem{healthnav}
HealthNav: Healthnav baycare health system, inc.
  \url{https://play.google.com/store/apps/details?id=baycare.com.healthnav}
  (2021)

\bibitem{IkramBK19}
Ikram, M., Beaume, P., Kaafar, M.A.: Dadidroid: An obfuscation resilient tool
  for detecting android malware via weighted directed call graph modelling. In:
  SECRYPT (2019)

\bibitem{AdBlocking}
Ikram, M., Kaafar, M.A.: A first look at mobile ad-blocking apps. In: 2017 IEEE
  16th International Symposium on Network Computing and Applications (NCA).
  pp.~1--8. IEEE (2017)

\bibitem{ikram_vtscore}
Ikram, M., Vallina-Rodriguez, N., Seneviratne, S., Kaafar, M.A., Paxson, V.: An
  analysis of the privacy and security risks of android vpn permission-enabled
  apps. In: IMC (2016)

\bibitem{SymDriod}
Jeon, J., Micinski, K.K.: Symdroid : Symbolic execution for dalvik (2013)

\bibitem{Leontiadis2012}
Leontiadis, I., Efstratiou, C., Picone, M., Mascolo, C.: Don’t kill my ads!
  balancing privacy in an ad-supported mobile application market.
  \doi{10.1145/2162081.2162084}

\bibitem{lindorfer2014andrubis}
Lindorfer, M., Neugschwandtner, M., Weichselbaum, L., Fratantonio, Y., {van der
  Veen}, V., Platzer, C.: {ANDRUBIS - 1,000,000 Apps Later: A View on Current
  Android Malware Behaviors}. In: BADGERS 2014 (2014)

\bibitem{gplaycli}
matlink: Google play downloader via command line v3.25.
  \url{https://github.com/matlink/gplaycli} (2018), accessed: 10/10/2020

\bibitem{Mediktor}
Mediktor: Mediktor tecker medical.
  \url{https://play.google.com/store/apps/details?id=com.teckelmedical.mediktor}
  (2021)

\bibitem{exported_hack}
Melamed, T.: Hacking android apps through exposed components.
  {https://www.linkedin.com/pulse/hacking-android-apps-through-exposed-components-tal-melamed}
  (2020)

\bibitem{mitmproxy}
mitmproxy: - an interactive https proxy. \url{https://mitmproxy.org} (2020)

\bibitem{apkid_diss3_adv}
Nair, R.: Techbliss - tutorial anti-disassembly techniques used by malware (a
  primer).
  {https://www.techbliss.org/threads/anti-disassembly-techniques-used-by-malware-a-primer-by-rahul-nair.804/}
  (2015)

\bibitem{keytool}
Oracle: keytool.
  \url{https://docs.oracle.com/javase/8/docs/technotes/tools/unix/keytool.html}
  (2020), accessed: 18/12/2020

\bibitem{apkid_anti-debug}
OWASP: Testing anti-debugging detection (mstg-resilience-2) - android
  anti-reversing defenses.
  \url{https://mobile-security.gitbook.io/mobile-security-testing-guide/android-testing-guide/0x05j-testing-resiliency-against-reverse-engineering}
  (2020), oWASP Mobile Security Guide - Accessed: 18/01/2020

\bibitem{zangfor_misc2}
Reddit, F.: Virustotal detects malware.
  \url{https://www.reddit.com/r/pathofdiablo/comments/ewooys/virustotal\_detects\_malware/}
  (2020)

\bibitem{apkid_rednaga}
RedNaga: Apkid - anti-analyisis open source tools.
  \url{https://github.com/rednaga/APKiD} (2016)

\bibitem{ren2018appversions}
Ren, J., Lindorfer, M., Dubois, D.J., Rao, A., Choffnes, D.R.,
  Vallina-Rodriguez, N.: Bug fixes, improvements, ... and privacy leaks: A
  longitudinal study of pii leaks across android app versions. In: NDSS (2018)

\bibitem{apkid_manipulator1}
Security, R.: Apkid and android compiler fingerprinting.
  \url{https://rednaga.io/2016/07/30/apkid_and_android_compiler_fingerprinting/}
  (2016)

\bibitem{PaidAppsTracking}
Seneviratne, S., Kolamunna, H., Seneviratne, A.: A measurement study of
  tracking in paid mobile applications. In: WiSeC (2015)

\bibitem{apkid_diss1}
SIMFORM: How to avoid reverse engineering of your android app?
  \url{https://www.simform.com/how-to-avoid-reverse-engineering-of-your-android-app/}
  (2015)

\bibitem{smahel2019functions}
Smahel, D., Elavsky, S., Machackova, H.: Functions of mhealth applications: A
  user’s perspective. Health informatics journal  \textbf{25}(3),  1065--1075
  (2019)

\bibitem{cert_sign2}
SSL: Why you need code signing certificate for your android app?
  \url{https://aboutssl.org/why-you-need-code-signing-certificate-for-android-app/}
  (2020)

\bibitem{sha-1_attack}
Stevens, M., Bursztein, E., Karpman, P., Albertini, A., Markov, Y.: {The First
  Collision for Full SHA-1}. In: {International Cryptology Conference ---
  CRYPTO 2017}. pp. 570--596. {IACR}, Santa-Barbara, United States (Aug 2017).
  \doi{10.1007/978-3-319-63688-7\_19},
  \url{https://hal.archives-ouvertes.fr/hal-01982005}

\bibitem{tangari2021mobile}
Tangari, G., Ikram, M., Ijaz, K., Kaafar, M.A., Berkovsky, S.: Mobile health
  and privacy: cross sectional study. bmj  \textbf{373} (2021)

\bibitem{trifan2019passive}
Trifan, A., Oliveira, M., Oliveira, J.L.: Passive sensing of health outcomes
  through smartphones: Systematic review of current solutions and possible
  limitations. JMIR mHealth and uHealth  (2019)

\bibitem{virustotal}
VirusTotal: Multitude anti-virus engines.
  \url{https://www.virustotal.com/gui/home/upload} (2020)

\bibitem{WebMD}
WebMD: Webmd: Check symptoms, rx.
  \url{https://play.google.com/store/apps/details?id=com.webmd.android} (2021)

\bibitem{apkid_obs1}
Weerasekara, S.I.: Gitconnected - obfuscation in android.
  \url{https://levelup.gitconnected.com/android-obfuscation-e608f79f0d09}
  (2020)

\bibitem{zhao2019decade}
Zhao, B.Z.H., Ikram, M., Asghar, H.J., Kaafar, M.A., Chaabane, A.,
  Thilakarathna, K.: A decade of mal-activity reporting: A retrospective
  analysis of internet malicious activity blacklists. In: Proceedings of the
  2019 ACM Asia Conference on Computer and Communications Security. pp.
  193--205 (2019)

\end{thebibliography}

\appendix

\begin{table*}
  \caption{Overview of our analyses of SymtomCheckers. \textbf{Col.2-4:} SymptomChecking apps IDs, number of installs, and average rating (Av. Rat.). \textbf{Col.5-8:} Certificate signing algorithm, number of exported components, VirusTotal's number of positives, number of anti-analysis techniques implemented by SymtomCheckers. \textbf{Col.9-11:} Fraction of requests communicated with HTTP, number of third-parties domains, and fraction of negative reviews to the total number of users' reviews.}
  \label{tab:apps_list}
  \tabcolsep=0.08cm
  \centering
  \small
  \scalebox{0.7} {
  \begin{tabular}{|l|l|r|r|l|r|r|r|r|r|r|}
    \hline
    \hline
        {\bf No.} & {\bf Apps Name}  & {\bf \# of Installs} & {\bf Av. Rat.} & {\bf Cert.} & {\bf \# of Exp.} & {\bf VTS}  & {\bf \# of Anti} & {\bf \% HTTP} & {\bf \# Third-}& {\bf \% of Neg.}\\ 
        
        {\bf} & {\bf } & {\bf }  & {\bf } & {\bf Signing} & {\bf Comp.} & {\bf }  & {\bf Analysis}& {\bf }& {\bf party} & {\bf Review}\\ 
    \hline
        1 & com.webmd.android & 10,000,000+ & 4.44 & SHA1 & 17 & 0 & 2 & n.a. & 9 & 17.9\\ 
        2 & com.ada.app & 5,000,000+ & 4.74 & SHA256 & 3 & 0 & 1 & 37.5 & 4 & 3.7\\ 
        3 & md.your & 1,000,000+ & 4.1 & SHA256 & 12 & 0 & 3 & 5.13 & 6 & 15.3\\ 
        4 & com.mayoclinic.patient & 1,000,000+ & 3.9 & SHA1 & 12 & 1 & 3 & 2.11 & 1 & 32.1\\ 
        5 & com.programming.progressive & & &  &  &  &  &  &  & \\
        & .diagnoseapp & 500,000+ & 4.52 & SHA256 & 0 & 0 & 1 & 25 & 2 & 5.5\\ 
        6 & com.symptomate.mobile & 100,000+ & 4.41 & SHA1 & 1 & 0 & 1 & 1.49 & 0 & 11\\ 
        7 & nl.japps.android.depressiontest & 100,000+ & 3.82 & SHA1 & 4 & 0 & 2 & 3.45 & 2 & 13.9\\ 
        8 & air.com.sensely.asknhs & 100,000+ & 4.29 & SHA256 & 2 & 0 & 2 & 1.80 & 2 & 25.7\\ 
        9 & com.caidr & 100,000+ & 4.08 & SHA256 & 16 & 0 & 2 & 1.92 & 10 & 33.6\\ 
        10 & com.teckelmedical.mediktor & 50,000+ & 3.62 & SHA256 & 7 & 1 & 2 & 2.08 & 9 & 30\\ 
        11 & au.org.healthdirect.healthdirectapp & 50,000+ & 3.88 & SHA256 & 8 & 0 & 2 & 2.15 & 5 & 25.4\\ 
        12 & ai.mayamd.mayamd & 50,000+ & 4.3 & SHA256 & 18 & 0 & 2 & 3.23 & 9 & 9.8\\ 
        13 & de.msmix.foody.android & 10,000+ & 4.11 & SHA256 & 1 & 0 & 2 & 37.50 & 1 & 13.3\\ 
        14 & dk.kalorieopslag.symdir & 10,000+ & 3.39 & SHA256 & 0 & 0 & 3 & 40 & 3 & 28.6\\ 
        15 & com.appcolliders.doctordiagnose & 10,000+ & 3.5 & SHA256 & 2 & 0 & 1 & 0 & 1 & 47.8\\ 
        16 & com.epainassist.symptomchecker & 10,000+ & 3.28 & SHA256 & 1 & 0 & 3 & 1.02 & 0 & 44.4\\ 
        17 & baycare.com.healthnav & 10,000+ & 4.18 & SHA256 & 4 & 1 & 2 & 7.89 & 8 & 34.5\\ 
        18 & com.app.babycheck & 10,000+ & 3.03 & SHA256 & 0 & 0 & 2 & 2.86 & 1 & 63.3\\ 
        19 & com.bluecreate.vitamincheck & 10,000+ & 4.42 & SHA256 & 16 & 0 & 2 & n.a. & 0 & 2.4\\ 
        20 & ai.tibot & 10,000+ & 3.74 & SHA256 & 9 & 0 & 3 & 3.06 & 8 & 27.6\\ 
        21 & andy.symptomes & 10,000+ & 3.79 & SHA256 & 1 & 0 & 2 & 25 & 1 & 0\\ 
        22 & com.appstronautstudios.heada &  &  &  &  &  &  &  &  & \\ 
        & chetracker & 5,000+ & 4.19 & SHA256 & 6 & 0 & 2 & 12.50 & 3 & 15.6\\ 
        23 & com.app.diagnose & 5,000+ & 2.8 & SHA256 & 6 & 0 & 2 & 1.41 & 2 & 38.5\\ 
        24 & com.ro.doctor31 & 5,000+ & 3.69 & SHA256 & 1 & 0 & 2 & 12.50 & 1 & 40\\ 
        25 & com.worthyworks.multihtml & 1,000+ & 4.55 & SHA256 & 0 & 0 & 1 & 2.48 & 2 & 5.3\\ 
        26 & com.sylextech.youshieldapp & 1,000+ & 4.09 & SHA256 & 0 & 0 & 2 & 20 & 0 & 14.3\\ 
        27 & developer007.magdy.symptomchecker & 1,000+ & n.a. & SHA256 & 0 & 0 & 1 & n.a. & 0 & n.a.\\ 
        28 & com.pharma.mydiagnosis & 1,000+ & 3.88 & SHA256 & 8 & 0 & 2 & 13.04 & 2 & 0\\ 
        29 & com.prembros.symptomator & 1,000+ & 3.64 & SHA256 & 4 & 0 & 1 & 33.33 & 0 & 0\\ 
        30 & com.myownmed.quinnipiac & 500+ & n.a. & SHA256 & 5 & 0 & 2 & n.a. & 3 & n.a.\\ 
        31 & com.francotiveron.SymptomsChecker & 500+ & n.a. & SHA256 & 0 & 0 & 0 & 7.89 & 0 & n.a.\\ 
        32 & com.worthyworks.myapplication & 500+ & 3.5 & SHA256 & 2 & 0 & 1 & n.a. & 0 & 20\\ 
        33 & com.dietchartapp.medweiser & 100+ & 3.67 & SHA256 & 0 & 0 & 0 & 0.32 & 0 & 50\\ 
        34 & com.nextgenblue.symptomchecker & 100+ & n.a. & SHA256 & 8 & 1 & 2 & n.a. & 0 & n.a.\\ 
        35 & com.socmedica.symptomchecker & 50+ & n.a. & SHA256 & 18 & 0 & 1 & 10 & 8 & n.a.\\ 
        36 & de.pr2.ebmapp & 10+ & n.a & SHA256 & 0 & 0 & 1 & 46.43 & 0 & n.a\\     
\hline\hline
  \end{tabular}
  }
\end{table*}

\end{document}